\documentclass[aps,prl,reprint,superscriptaddress,citeautoscript,floatfix,longbibliography]{revtex4-1}
\usepackage{graphicx}
\usepackage{epstopdf}
\usepackage{hyperref}

\usepackage{float}
\usepackage{amsmath}
\usepackage[export]{adjustbox}
\usepackage{xcolor}
\usepackage{cancel}
\usepackage{textcomp}
\usepackage{gensymb}
\usepackage{multirow}
\usepackage{array}

\usepackage[pagewise]{lineno}

\usepackage[sort&compress]{natbib}

\usepackage[compact]{titlesec}
    \titlespacing{\section}{0pt}{2ex}{1ex}
    \titlespacing{\subsection}{0pt}{1ex}{1ex}
    \titlespacing{\subsubsection}{0pt}{0.5ex}{1ex}
\begin{document}


\title{High Frequency Nonlinear Response of Superconducting Cavity-Grade Nb surfaces}


\author{Bakhrom Oripov}
\email[Corresponding author:]{bakhromtjk@gmail.com}
\affiliation{Center for Nanophysics and Advanced Materials, Department of Physics, University of Maryland, College Park, MD 20742, USA}

\author{Thomas Bieler}
\affiliation{Chemical Engineering and Materials Science, Michigan State University, East Lansing, Michigan 48824, USA}

\author{Gianluigi Ciovati}
\affiliation{Thomas Jefferson National Accelerator Facility, Newport News, Virginia 23606, USA}

\author{Sergio Calatroni}
\affiliation{European Organization for Nuclear Research (CERN), 1211 Geneva 23, Switzerland}

\author{Pashupati Dhakal}
\affiliation{Thomas Jefferson National Accelerator Facility, Newport News, Virginia 23606, USA}

\author{Tobias Junginger}
\affiliation{Engineering Department, Lancaster University, Lancaster LA1 4YW, United Kingdom}

\author{Oleg B. Malyshev}
\affiliation{ASTeC, STFC Daresbury Laboratory, Warrington WA4 4AD, United Kingdom}

\author{Giovanni Terenziani}
\affiliation{European Organization for Nuclear Research (CERN), 1211 Geneva 23, Switzerland}

\author{Anne-Marie Valente-Feliciano}
\affiliation{Thomas Jefferson National Accelerator Facility, Newport News, Virginia 23606, USA}

\author{Reza Valizadeh}
\affiliation{ASTeC, STFC Daresbury Laboratory, Warrington WA4 4AD, United Kingdom}

\author{Stuart Wilde}
\affiliation{ASTeC, STFC Daresbury Laboratory, Warrington WA4 4AD, United Kingdom}

\author{Steven M. Anlage}
\affiliation{Center for Nanophysics and Advanced Materials, Department of Physics, University of Maryland, College Park, MD 20742, USA}


\date{\today}

\begin{abstract}
Nb superconducting radio-frequency (SRF) cavities are observed to break down and lose their high-Q superconducting properties at accelerating gradients below the limits imposed by theory. The microscopic origins of SRF cavity breakdown are still a matter of some debate. To investigate these microscopic issues, temperature- and power-dependent local third-harmonic response is measured
on bulk Nb and Nb thin-film samples using a novel near-field magnetic microwave microscope between 2.9 and 10 K and 2 and 6 GHz. Both periodic and nonperiodic response as a function of applied rf field amplitude are observed. We attribute these features to extrinsic and intrinsic nonlinear responses of the sample. The rf-current-biased resistively shunted junction (RSJ) model can account for the periodic response and fits very well to the data using reasonable parameters. The
nonperiodic response is consistent with vortex semiloops penetrating into the bulk of the sample once sufficiently high rf magnetic field is applied and the data can be fit to a time-dependent Ginzburg-Landau (TDGL) model of this process. The fact that these responses are measured on a wide variety of Nb samples suggests that we are capturing the generic nonlinear response of air-exposed Nb surfaces.
\end{abstract}


\maketitle


\section{I. INTRODUCTION}
Researchers studying high-energy physics are considering several next-generation accelerator designs, including the International Linear Collider (ILC), where a very large number (approximately 16000) of Nb superconducting radio-frequency (SRF) cavities will be employed \cite{ILCref3,ILC}. Each cell works under high rf fields with the highest electric field on the accelerating axis and the maximum magnetic field on the equator of the cylindrical cavity \cite{aune_superconducting_2000}. \par 
The high-accelerating gradient performance of Nb SRF cavities is often limited by breakdown events below the intrinsic limiting surface fields of Nb. These breakdowns are often caused by defects at discrete locations inside the cavity \cite{ge_routine_2011, iwashita_development_2008}. Although cavities with high-quality surfaces with excellent rf properties had been achieved, there is a lack of detailed understanding of the causal links among surface treatments, defects, and ultimate rf performance at low temperatures with many theoretical models being proposed to address the issue \cite{gurevichMultiscale,gurevich_dynamics_2008,RomanenkoLowTLS,Gurevich_RFtheory}. Cavity uniformity will be one of the key requirements of the ILC; thus, the ability to mass produce high-quality SRF cavities and diagnose them rapidly is very important.  \par
Lately, various optical and other inspection techniques for finished cavities have been developed. For example a temperature map of the cavity exterior is taken and hot spots are identified \cite{knobloch_design_1994, canabal_development_2007}. Then a high-resolution optical image of the cavity interior is taken and pits and other defects are identified \cite{iwashita_development_2008, watanabe_review_nodate}. Later, the correlation between defects and hot spots is studied. As an example, a theory that describes the relation between the breakdown field and the geometry of surface pits has been developed \cite{KuboPit}. Currently, studies that focus on elemental composition and purity of material \cite{safa_advances_1996}, dc critical magnetic field or superheating field \cite{malyshev_first_2016}, postmortem microanalysis of hot-cold spots \cite{GigiBielerPits},\cite{trenikhina_nanostructural_2015} or various sophisticated optical inspection tools \cite{iwashita_development_2008, watanabe_review_nodate} make up the backbone of SRF material science. These efforts have resulted in advancements in cavity treatment recipes that have led to SRF cavities with high gradients and high-quality factors $ > 3 \times {10}^{10}$ at 1.3 GHz and 2K. However, it is not possible to locally and directly probe rf performance of the active interior surface of a cavity at low temperatures. \par
Finally, there is also interest in looking beyond Nb to superconductors like $MgB_2$ and other superconductor coatings or the possibility of reducing the cost by using Nb-coated copper cavities. There is also a proposal to create superconductor/insulator multilayer thin-film coatings with enhanced rf critical fields \cite{gurevich_enhancement_2006}. However, often it is very costly and/or difficult to build full-size SRF cavities with high-quality versions of these new materials to see if they really are superior.  Hence there is a need to quantitatively examine these materials at high frequencies and low temperatures using simpler methods that effectively reproduce the demanding conditions found operating in SRF cavities. \par
Based on the needs of the SRF community, we successfully build a near-field magnetic microwave microscope using a magnetic writer from a conventional magnetic recording hard-disk drive \cite{lee_microwave_2005,mircea_phase-sensitive_2009,tai_nonlinear_2011,tai_nanoscale_2013,tai_near-field_2014,tai_nanoscale_2015}. This microscope is capable of locally measuring both linear and nonlinear rf response. The nonlinear response is far more sensitive to applied rf power than the linear response \cite{oates_intermodulation_2007} and it can be linked to the nonlinear surface impedance of the sample \cite{oates_overview_2007}, which is an important parameter for a SRF cavity driven by high rf magnetic field. Furthermore, surface defects are expected to locally suppress the rf critical current, which leads to a higher nonlinear response, thus making this microscope an excellent defect detector \cite{lee_doping-dependent_2005, booth_-wafer_2001}.

\begin{figure}
    \centering
    \includegraphics[width=0.45\textwidth]{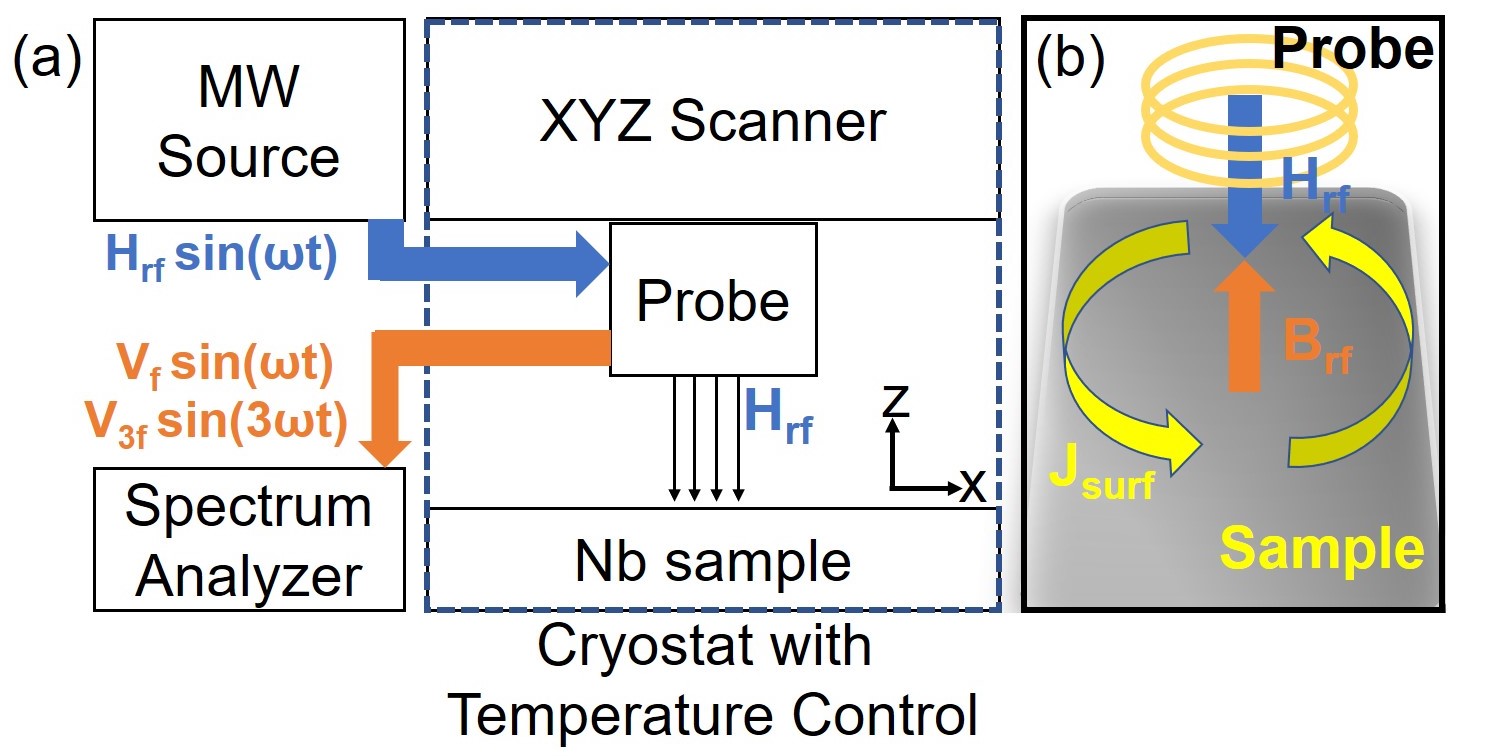}
    \caption{ (a) Schematic of experimental setup. A microwave (MW) source is used to generate rf signal and feed it to the magnetic probe. The sample response magnetic field is coupled back to the probe and measured with a spectrum analyzer. (b) Sketch of the probe-sample interaction. The magnetic probe is approximated as a current loop producing perpendicular magnetic field inducing screening currents on the surface of the sample. This current generates a response magnetic field that is coupled back to the same probe.}
    \label{ExperimentalSetup}
\end{figure}

\section{II. EXPERIMENTAL SETUP}
The magnetic write head inside a conventional hard disk-drive is a true engineering marvel. Nearly 1000 processing steps are needed to fabricate one \cite{APSwriter_pres}, but once finished a write head can produce $B_{RF}\approx600$-$mT$ rf magnetic field localized to an approximately $100$-$nm$ length scale \cite{SeagateInfo}. In this experiment, a Seagate perpendicular magnetic writer head is attached to a cryogenic XYZ positioner and used in a scanning probe fashion. Electrical contacts are made by directly soldering SMA coaxial connectors to the pads on the end of the writer probe transmission line (see Fig. S1 in the Supplemental Material \cite{Supp,TinkhamBCSGap,cyrot_ginzburg-landau_1973,aranson_dynamics_2001,Sorensen,Vinokur,TaminThesis}). Probe characterization results and other details can be found in \cite{tai_nonlinear_2011, tai_nanoscale_2013, tai_near-field_2014, tai_nanoscale_2015}. A microwave source is connected to the probe by coaxial cables (Fig. \ref{ExperimentalSetup}(a)). The probe is in contact with the sample, which is located on the cold plate of an Entropy 4K L-series cryostat with a base temperature of 2.8K. The probe produces a rf magnetic field perpendicular to the sample surface. The sample is in the superconducting state, so to maintain the Meissner state, a screening current is induced on the surface(Fig. \ref{ExperimentalSetup}(b)). This current generates a response magnetic field that is coupled back to the same probe, creates a propagating signal on the attached transmission line structure, and is measured with a spectrum analyzer at room temperature. Since superconductors are intrinsically nonlinear \cite{xu_nonlinear_1995}, both linear and nonlinear responses to an applied rf magnetic field are expected. To improve the signal-to-noise ratio, stray nonlinear signals produced by the amplifiers and the spectrum analyzer itself have to be suppressed. For example, high pass filters are installed between the probe and spectrum analyzer in order to block the fundamental input frequency signal from reaching the spectrum analyzer and producing nonlinear signals \cite{lee_microwave_2005,mircea_phase-sensitive_2009,tai_nonlinear_2011,tai_nanoscale_2013,tai_near-field_2014,tai_nanoscale_2015}. Measurements are performed at a fixed input frequency while the temperature of the sample and applied rf field amplitude are varied. No external dc magnetic field is applied. The entire setup is shielded from external magnetic field by a cylindrical superconducting shield installed around the coldest enclosure of the cryostat. \par
The present microscope is an upgrade of an older-generation setup reported in Refs. \cite{tai_nonlinear_2011, tai_nanoscale_2013, tai_near-field_2014, tai_nanoscale_2015}. The newer-generation magnetic write heads are able to induce stronger and more localized rf magnetic fields on the sample. A new cryogen-free cryostat enables collection of more data and detailed examination of a wide variety of bulk and thin- film Nb samples from different sample growers.

\section{III. DATA}
First, a bulk Nb sample provided by the Bieler group (MSU) originating from a Tokyo-Denkai large-grain Nb single crystal is examined. The surface of the sample is mechanically polished and then electropolished. Later, the sample is strained to $40\%$ elongation, cut in half, and two differently oriented halves (F and C) are welded back together \cite{tai_nanoscale_2015, DogBoneSample}. The sample is then measured for a third-harmonic response at a fixed position on side C of the sample 4 mm away from the weld. This location experiences a brief thermal excursion that retains much of the deformed dislocation defect structure as described in more detail in Fig. 16 of Ref. \cite{DogBoneSample} and Ref. \cite{DerekThesis}. \par

\begin{figure}
    \centering
    \includegraphics[width=0.45\textwidth]{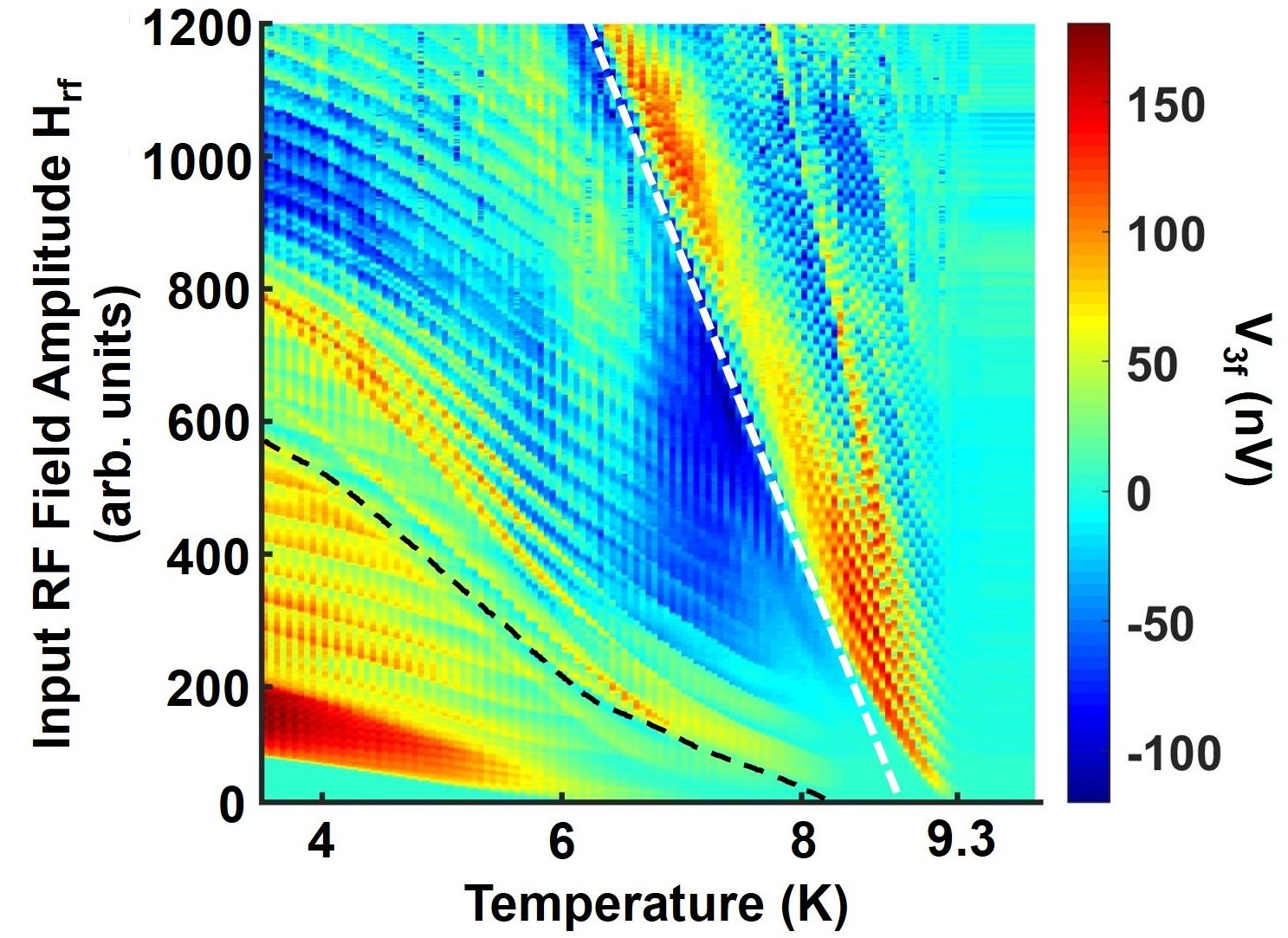}
    \caption{(Color online) Third-harmonic voltage $V_{3f}$ data vs temperature and applied rf field amplitude from a large-grain bulk Nb measured at a $4.38$-$GHz$ rf input frequency. A temperature-independent probe background nonlinearity has been subtracted, resulting in some negative values of $V_{3f}$.}
    \label{BulkNbSurf}
\end{figure}

Fig. \ref{BulkNbSurf} shows a plot of measured third-harmonic response $(V_{3f})$ vs temperature $(T)$ and applied rf field amplitude $(H_{rf})$ at an$f=4.38$-$GHz$ input frequency. A temperature-independent probe background is subtracted from the data for clarity. One notes several families of nonlinear response (separated by white and black dashed lines) that are roughly periodic as a function of applied rf amplitude.

\begin{figure}
    \centering
    \includegraphics[width=0.45\textwidth]{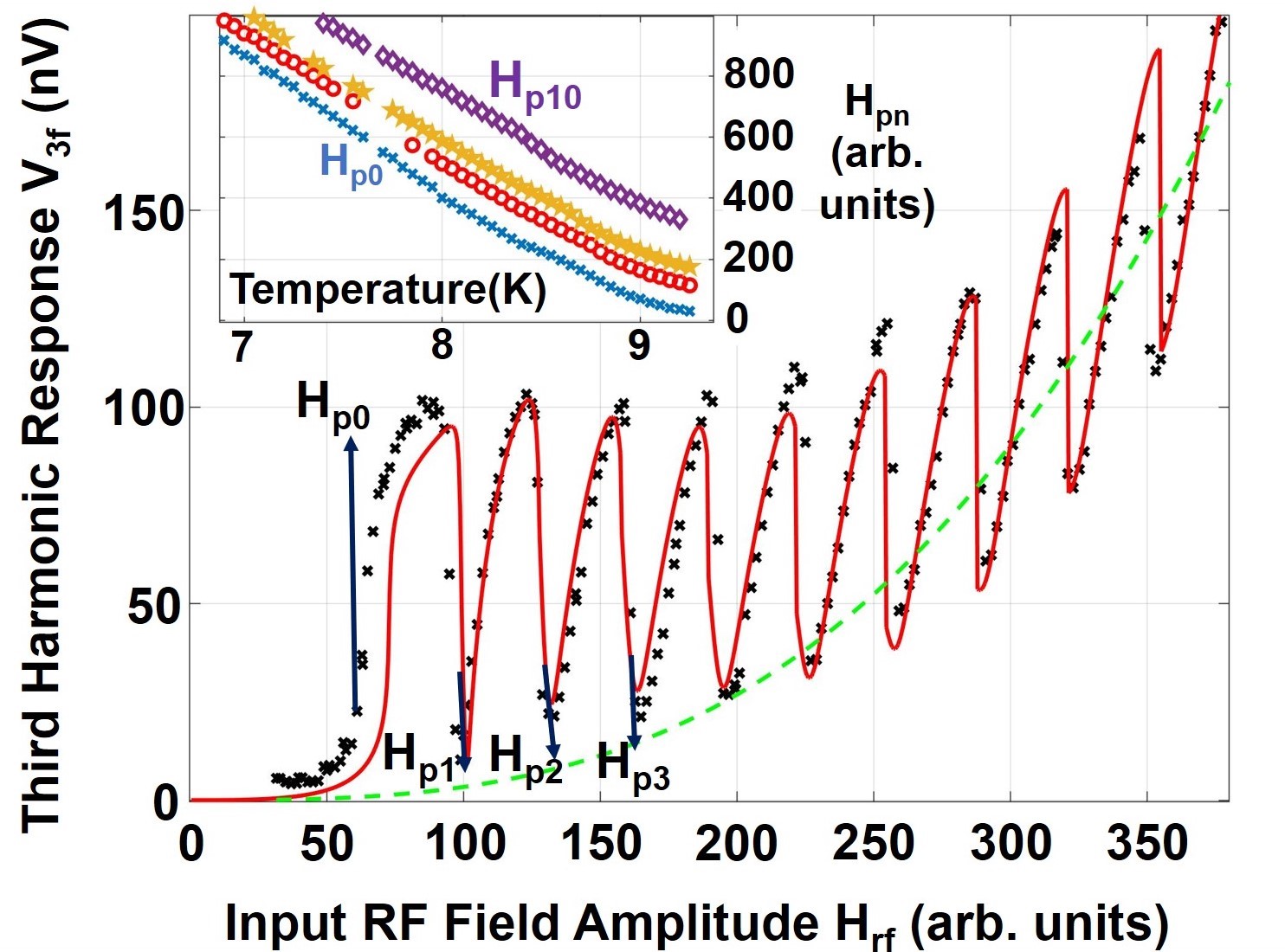}
    \caption{(Color online) Bulk Nb data at T=9.0K (black $\times$ markers), probe background (green dashed line) and RSJ fit with $I_{c}R=49.4\mu V$ (red solid line). Inset: $H_{p0}$ (blue $\times$), $H_{p3}$ (red circle), $H_{p5}$ (orange $\star$) and $H_{p10}$ (purple $\diamond$) vs temperature (K) from the data shown in Fig. \ref{BulkNbSurf}.}
    \label{BulkNbLine}
\end{figure}

Fig. \ref{BulkNbLine} shows a vertical line cut through this image ($V_{3f}$ vs $H_{rf}$) at a constant temperature of $T=9.0K$. As shown in the figure, $H_{p1}$, $H_{p2}$ and $H_{p3}$ are the rf field amplitudes corresponding to the first three dips in $V_{3f}(H_{rf})$. At all temperatures, $V_{3f}$ is periodic as a function of $H_{rf}$ above a temperature-dependent onset amplitude $H_{p0}(T)$. There is no temperature-dependent $V_{3f}$ signal above $9.3K$, indicating that the superconducting sample is indeed the source of the signal. To compare the temperature dependence of the dips, a temperature-dependent "period" and onset of periodicity $H_{p0}(T)$ are defined. At a constant temperature $T$, all dips $H_{pn}(T)$, where $n$ stands for the $n$'th dip, are fit to a straight line as a function of $n$. The slope of the line is defined as "period" and the $y$-intercept is defined as the onset of periodicity $H_{p0}(T)$ (Fig. S3 in \cite{Supp}). The inset of Fig. \ref{BulkNbLine} shows the temperature dependence of $H_{p0}(T)$, $H_{p3}(T)$, $H_{p5}(T)$ and $H_{p10}(T)$. Fig. \ref{BulkNbSurf} shows several families of similar features, each with a different onset of periodicity $H_{p0}$, period and onset temperature, suggesting that several nonlinear sources are within the field of view of the probe. \par

\begin{figure}
    \centering
    \includegraphics[width=0.45\textwidth]{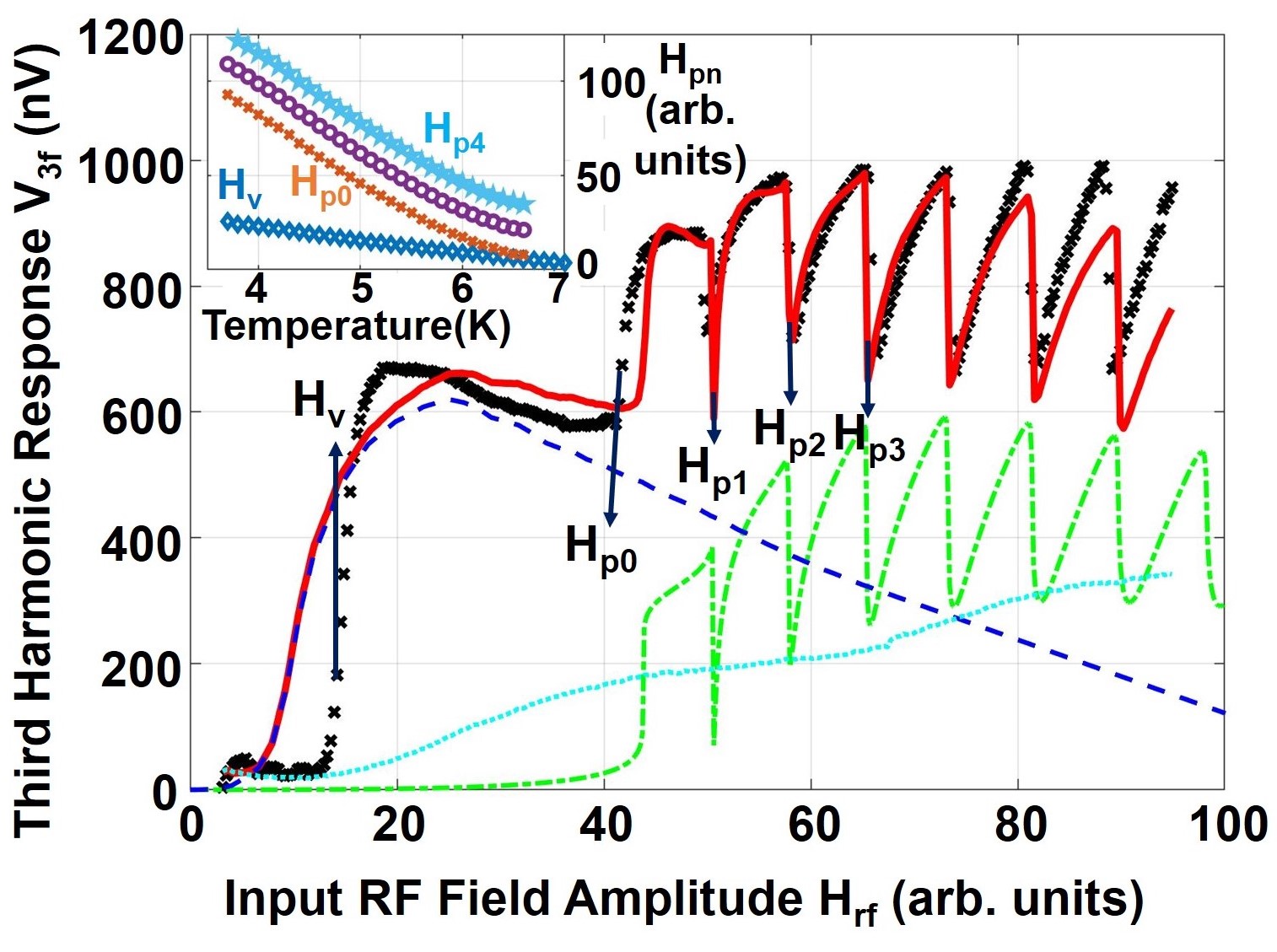}
    \caption{(Color online) Third-harmonic response $V_{3f}$ data at T=5.1K and 2.2 GHz on a Nb on Cu film sample from CERN (black $\times$ markers). Also shown are the RSJ fit with $I_{c}R=58.3\mu V$ (green dashed-dotted line), time-dependent Ginzburg-Landau fit with $H_{dp}=12$ (blue dashed line),  probe background (cyan dotted line) and complete fit obtained by vector complex addition (red solid line). Inset: $H_{p0}$ (red $\times$), $H_{p2}$ (purple circle), $H_{p4}$ (cyan $\star$) and onset of nonlinearity $H_{v}$ (blue $\diamond$) vs temperature (K).}
    \label{FilmNbLine}
\end{figure}

Also, a thin-film sample is prepared at CERN by depositing Nb on a copper substrate by high-power impulse magnetron sputtering (HIPIMS). The plasma is driven with rectangular voltage pulses in a frequency range of $50-500 Hz$ and pulse duration range of $50-200\mu s$. The peak discharge current
density is kept in the range of $0.2-1 A/{cm}^2$. This sample is well characterized by point-contact tunneling spectroscopy, low-energy muon spin rotation, and SEM. The point-contact tunneling measurements revealed that the sample has a distribution of superconducting gap values with a peak at $\Delta=1.48 meV$ \cite{junginger_low_2017}. However, the tunneling data also show a few points with large zero-bias conductance peaks. The sample has a fine-grain structure with $50$-$100$-$nm$ average grain size \cite{g_terenziani_niobium_2013}. Fig. \ref{FilmNbLine} shows the measured $V_{3f}$ vs $H_{rf}$ at $T=5.1 K$ and $2.2\ GHz$ rf input frequency. Similar to the bulk Nb sample, it also has a periodic response as a function of driving rf amplitude. However, the onset of periodicity $H_{p0}$ is different from the onset of nonlinear response $H_{v}$. The inset shows the temperature dependence of $H_{p0}(T)$, $H_{p2}(T)$, $H_{p4}(T)$, and $H_{v}(T)$. Clearly, $H_{p0}$ and $H_{v}$ have different temperature dependencies, but the $H_{pn}(T)$ temperature dependencies are similar to those seen on the bulk Nb samples.\par

Measurements are performed on several other bulk Nb and Nb film samples and a very similar periodic response of $V_{3f}$ vs $H_{rf}$ is observed, showing that this is a generic response of this material (Figs. \ref{2ndBulkNb}-\ref{AnneMarie} in the Appendix). Overall, two types of nonlinear response are observed: a low-field nonperiodic and a higher-field periodic response. All Nb samples show either periodic or low-field response, or both, depending on the spatial location of the magnetic writer probe on the surface of the sample.

\section{IV. MODELS}
The oxidation of Nb when exposed to air is a well-known and complex phenomenon \cite{halbritter_oxidation_1987,Halbritter2,AntoineTomography,NICO_NbOxides}. Oxygen forms a solid solution in Nb and produces materials with a continuous range of transition temperatures below the bulk $T_c$ of pure Nb. A thin Nb-oxide layer on the surface of Nb sandwiched with a Nb superconductor on both sides can create a superconductor-insulator-superconductor (S-I-S) or superconductor-normal-metal-superconductor (S-N-S) Josephson junction \cite{tai_nanoscale_2015}. Many theoretical works modeling linear and nonlinear rf impedance of weak-link Josephson junctions were previously published \cite{PortisBook, XieRCSJ, mcdonald_microwave_1997, ZhaiLongJunction, GurevichLongJunction}. Experimental data showing the rf response of a weak-link Josephson junction in a high-temperature superconductor is also available \cite{PortisBook,OatesJunctionVortex,lee_microwave_2005}.  \par
In our experiment, the probe induces rf current on the surface of the sample. This current can bias a junction at or near the surface. The Josephson junction ($JJ$) is modeled as an ideal short junction shunted by a resistance ($R$) and a capacitance ($C$) to form a parallel circuit \cite{mcdonald_microwave_1997}. For a small driving frequency compared to the plasma frequency $\omega_{p}=\sqrt{\cfrac{2\pi I_c}{\Phi_0C}}$ of the junction, the model can be simplified and the capacitive branch can be ignored. Here $I_c$ is the critical current of the junction and $\Phi_0$ is the magnetic flux quantum. The simplified circuit equation becomes $I_c \sin \delta+\cfrac{\Phi_0}{2\pi R}\cfrac{\partial \delta}{\partial t} = I_\omega \sin (\omega t)$, where $\delta$ is the gauge-invariant phase difference on the JJ, $V = \cfrac{\Phi_0}{2\pi}\cfrac{d \delta}{d t}$ is the potential difference across the junction and $I_\omega \sin (\omega t)$ is the rf current bias. In general, the voltage drop on the junction is nonsinusoidal in time and contains harmonics and subharmonics of the driving frequency $\omega$. One can calculate third-harmonic voltage $V_{3\omega}$ across the junction as a function of driving current amplitude $I_{\omega}$ for a given critical voltage of the junction $I_{c}R$. The result closely resembles the experimental data (see the red solid line in Fig. \ref{BulkNbLine}, green dashed-dotted line in Fig. \ref{FilmNbLine}, and Fig. S4 in \cite{Supp}). Each subsequent peak in $V_{3\omega}$ corresponds to an additional $2\pi$ phase slip across the junction in each rf cycle. \par

For the current-biased resistively shunted junction (RSJ) model, the onset-to-period ratio uniquely and monotonically depends on $I_{c}R$ (Fig. S5 inset in \cite{Supp}). Thus the measured onset-to-period ratio from the data taken at a constant temperature allows direct extraction of $I_{c}R(T)$ for the junction. In turn $I_{c}R(T)$ can be used to calculate the superconducting gap $\Delta(T)$, and $\Delta(0)$ can be obtained from a BCS temperature dependence fit to $\Delta(T)$ (Fig. S6 in \cite{Supp}). Note that the onset-to-period ratio provides a direct link between experimental data and numerical calculations from the model, and that the measured voltages are across the pads of the writer head, whereas the calculated voltages are across the junction. Thus, the RSJ fit plots in both Fig. \ref{BulkNbLine} and Fig. \ref{FilmNbLine} involve unknown scaling factors for both the rf amplitude and third-harmonic response. \par

\begin{table}
\begin{ruledtabular}
\begin{tabular}{| c | c | c | c |}
 Sample & \multicolumn{1}{m{2cm}|}{Weak-link Family} & $\Delta(0)$ $(meV)$ & $T_c (K)$ \\
 \hline
 \multirow{3}{6em}{Bulk Nb from MSU} & 1 & 0.80 & 9.28 \\ 
 & 2 & 0.35 & 8.41 \\
 & 3 & 0.33 & 7.28 \\
 \hline
 \multicolumn{1}{|m{2cm}|}{Nb on Cu from CERN} & 1 & 0.25 & 6.76 \\ [0.5ex] 
\end{tabular}
\end{ruledtabular}
\caption{Summary of fit superconducting gap and $T_c$ values obtained from $I_cR(T)$ extracted from third-harmonic data on two Nb samples. Weak-link number indicates the family number in the data. \label{GapTable}}
\end{table}
Superconducting gap values obtained from fits shown in Table.\ref{GapTable} are lower than the bulk value $\Delta(0)= 1.55 meV$ reported elsewhere \cite{proslier_tunneling_2008}. Nb samples with higher oxygen content tend to have lower superconducting gap values \cite{halbritter_oxidation_1987, sheet_possible_2010}. Thus the superconducting gap value at a $Nb/NbO_x/Nb$ weak-link can be lower due to a gradient in oxygen content.\par
Note that three families of periodic features are visible in Fig. \ref{BulkNbSurf}, with different periods and onsets $H_{p0}$. Table.\ref{GapTable} summarizes fits to these features as well as those on a thin-film sample. The fact that the responses from the weak-links have different superconducting gap $\Delta(0)$ values and $T_c$'s shows that they likely have different oxygen content.

\begin{figure}
    \centering
    \includegraphics[width=0.45\textwidth]{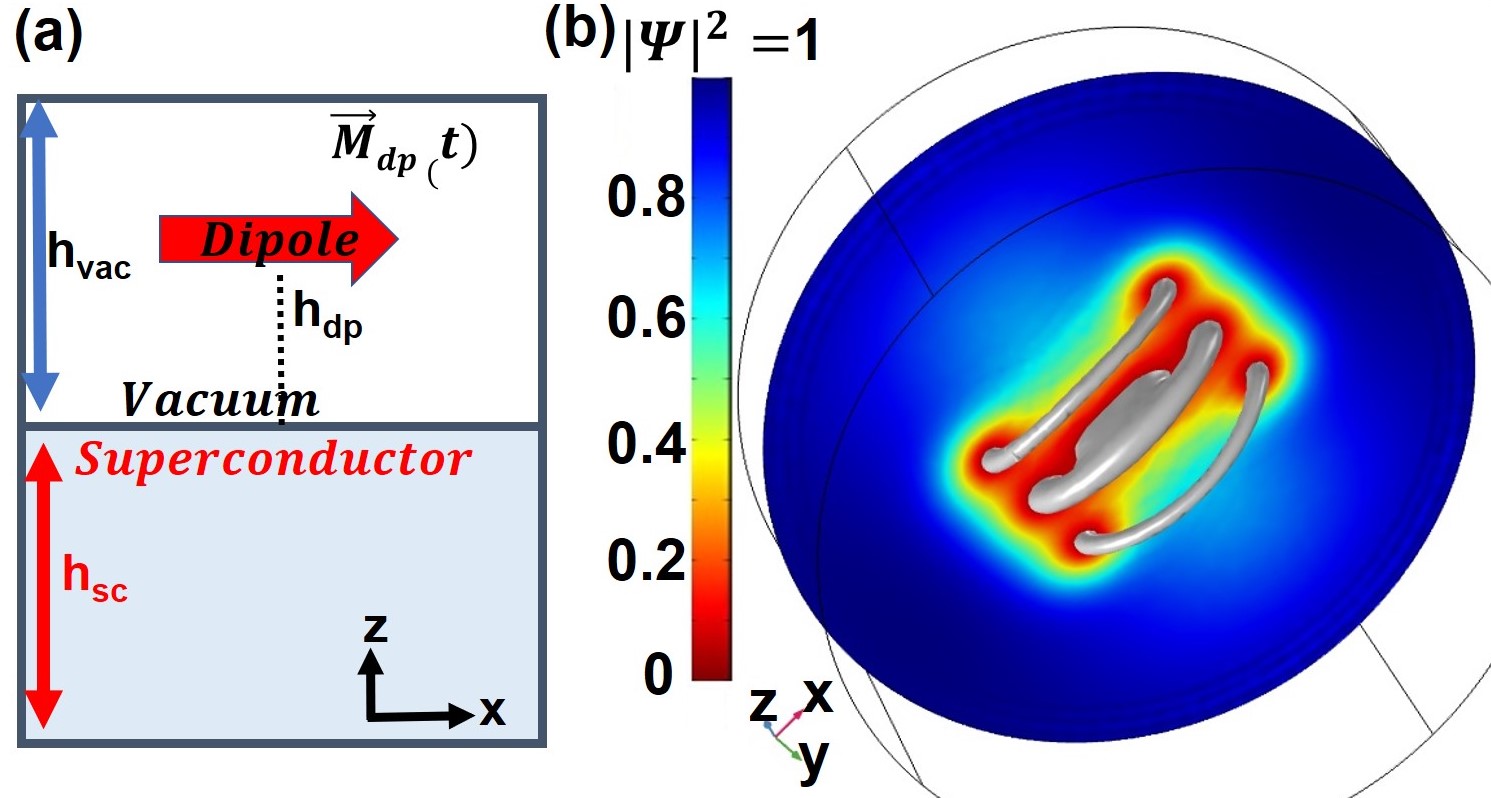}
    \caption{(Color online) (a) Schematic TDGL simulation setup. The magnetic probe is approximated as a point magnetic dipole parallel to the surface. The simulation is divided into two domains: superconductor and vacuum. (b) Superconducting order parameter $\vert \Psi \vert^2$ evaluated at the superconductor surface at time $t=73 \tau_0$ looking from inside the superconducting domain into the vacuum domain. The dipole magnetic moment $\vec{M}_{dp}(t)$ is chosen such that the applied rf magnetic field at the surface of the superconductor below the dipole $\vec{B}(t)=0.75 sin(\omega t) \hat{x}$. Grey contours represent surfaces of $\vert \Psi \vert^2=5 \times 10^{-3}$ and show three vortex semiloops penetrating inside the superconductor.} 
    \label{Comsol}
\end{figure}

Even in the absence of defects, a superconductor subjected to a strong rf magnetic field can generate a nonlinear rf response \cite{gurevich_dynamics_2008}. This scenario is modeled by solving the time-dependent Ginzburg-Landau (TDGL) equations with COMSOL simulation software. The magnetic writer probe was approximated to be a pointlike magnetic dipole with a sinusoidal time-dependent magnetic moment $\vec{M}_{dp}(t)$, located at a height $h_{dp}$ above the origin and parallel to the surface. The simulation is divided into two domains: superconductor: $-h_{sc}<z<0$, and vacuum $0<z<h_{vac}$ (see Fig.\ref{Comsol}.(a)). The TDGL equations are solved in the $1^{st}$ domain while only Maxwell's equations are solved in the second domain with appropriate boundary conditions at the interface. \par

Fig. \ref{Comsol}.(b) shows a snapshot of a typical TDGL simulation. As $\vec{M}_{dp}(t)$ of the dipole increases, a domain with a suppressed order parameter forms in the superconductor in a region below the dipole. Later in the rf cycle, as the magnetic field peaks and begins to decrease in magnitude the domain vanishes and vortex semiloops emerge (Fig. S9 in \cite{Supp}). The response rf magnetic field is calculated from the time-dependent screening currents and the third-harmonic response recovered at the location of the dipole is obtained through Fourier transformation. \par
The TDGL-derived third-harmonic voltage $V_{3\omega}$ as a function of the peak magnetic moment of the dipole shows the following behavior. $V_{3\omega}$ is small for small moments but increases rapidly above an onset magnetic moment value corresponding to an onset rf field amplitude $H_v$ at the surface, reaches a peak value and then slowly decreases back to a small finite value (see the blue long-dashed line in Fig.\ref{FilmNbLine}). Note that this evolution is very similar to the low-field response, an example of which is shown in Fig.\ref{FilmNbLine} (the $H_{rf} < H_{p0}$ part of the data).  \par

In total the data shown in Fig.\ref{FilmNbLine} have contributions from a nonlinear response generated by the current-biased junction, intrinsic low-field nonlinearity modeled by TDGL, and a temperature-independent probe background. First the onset-to-period ratio is calculated from the data. Then the $I_cR$ value corresponding to the onset-to-period ratio of the data is found (Fig. S5 in \cite{Supp}). Next, the horizontal axis scaling factor for the RSJ model is fixed to match the $H_{p0}$ from the RSJ model to the $H_{p0}$ from the data. Finally, complex vector addition of all three contributions is performed for a full fit (Fig. \ref{FilmNbLine}). The probe background was measured at 10K, which is above the $T_c$ of Nb. The phase of the probe background was assumed to be constant and is set to zero, while the complex values of the model responses are included in the sum. The horizontal and vertical axis scaling factors for the TDGL model and the vertical axis scaling factor for the RSJ model are the three fitting parameters.  \par

An estimate of the applied rf field strength created by the probe on the surface of the sample can be made as follows. At higher temperatures as $T \to T_c$, the upper critical magnetic field is given by $B_{c2}(T)\approx2B_{c2}(0)(1-T/T_c)$. By determining the $H_{rf}$ value at which the periodic nonlinear response is suppressed, and assuming $B_{c2}(0)=240mT$ \cite{aune_superconducting_2000,AnneMarieThesis} and $T_c=9.3K$, we can estimate the rf field experienced by the defect to be $11.4\mu T /(arb. unit)$. This approximately calibrates the vertical axis in Fig. \ref{BulkNbSurf} and the horizontal axis in Fig. \ref{BulkNbLine}. Similarly, using the horizontal scaling factor of $<X_{scale}>=0.22\mu V/(arb.units)$ obtained in the Supplemental Material \cite{Supp} for weak-link 1 in Fig. \ref{BulkNbSurf} and an estimated weak-link normal state resistance of $R_n=100 \Omega$ \cite{tai_nanoscale_2015}, we can estimate the driving current flowing through the weak-link to be $2.2 nA/(arb. unit)$. Thus, for weak-link 1 in Fig. \ref{BulkNbSurf}, this would correspond to a maximum rf magnetic field amplitude of $13.7mT$ and maximum driving current of $I_\omega=2.6\mu A$ with a zero-temperature critical current of $I_c(0)=12.6 \mu A$. \par

\section{V. DISCUSSION}
We measure a variety of Nb samples intended for SRF applications from five different sources (see the Appendix). A similar nonlinear response is observed in all of them. The intrinsic low-field nonlinear response calculated based on the Ginzburg–Landau (GL) model near
$T_c$ has a strong dependence on the sample thickness \cite{tai_nanoscale_2013,lee_doping-dependent_2005}. Therefore, bulk Nb samples tend to have a weaker intrinsic nonlinear response compared to thin-film Nb samples. Depending on the spatial location of the probe, we observe a nonlinear response, which can be considered as a low-field response, a periodic response, or a combination of both.  The response is also reproducible and the same result is obtained on a bulk large-grain Nb sample after an interval of 2 weeks, for example. \par
Note that the magnetic field created by the magnetic writer probe is very "aggressive" and introduces a perpendicular component; hence, it does not precisely replicate the parallel rf magnetic fields experienced inside a SRF cavity. Nevertheless, this method is a very useful tool to compare rf responses across a variety of samples and to identify the source of the nonlinearity. We can locate the weak-link, measure its superconducting gap, and compare local rf vortex nucleation fields $H_v$ on the surface of superconducting samples. \par

In summary, we observe two generic types of local nonlinear response on the surfaces of air-exposed Nb bulk and thin-films. The nonlinear response is quantitatively modeled and understood. The existence of weak-links on air-exposed SRF grade Nb samples is experimentally confirmed. We also demonstrate the ability to successfully differentiate between the sources of local nonlinearity based on their dependence on rf field amplitude and temperature.

\section{ACKNOWLEDGMENTS}
\begin{acknowledgments}
The authors would like to thank Michael Conover from Seagate Technology for providing magnetic write heads. This research is conducted with support from the U.S. Department of Energy/ High Energy Physics through Grant No.DESC0017931 (B.O.,S.M.A.). T.B. acknowledges support from the U.S. Department of Energy/Office of High Energy Physics contract DE-SC0009962. The work at Jefferson Lab (G.C., P.D., A.-M. V.-F.) is supported by Jefferson Science Associates, LLC, under U.S. DOE Contract No. DE-AC05-06OR23177. R.V. and S.W.'s contribution is conducted under the aegis of the Science and Technology Facility Council (STFC).
\end{acknowledgments}

\section{APPENDIX: DATA FROM OTHER NB SAMPLES \label{Appendix}}
\begin{figure}
    \centering
    \includegraphics[width=0.45\textwidth]{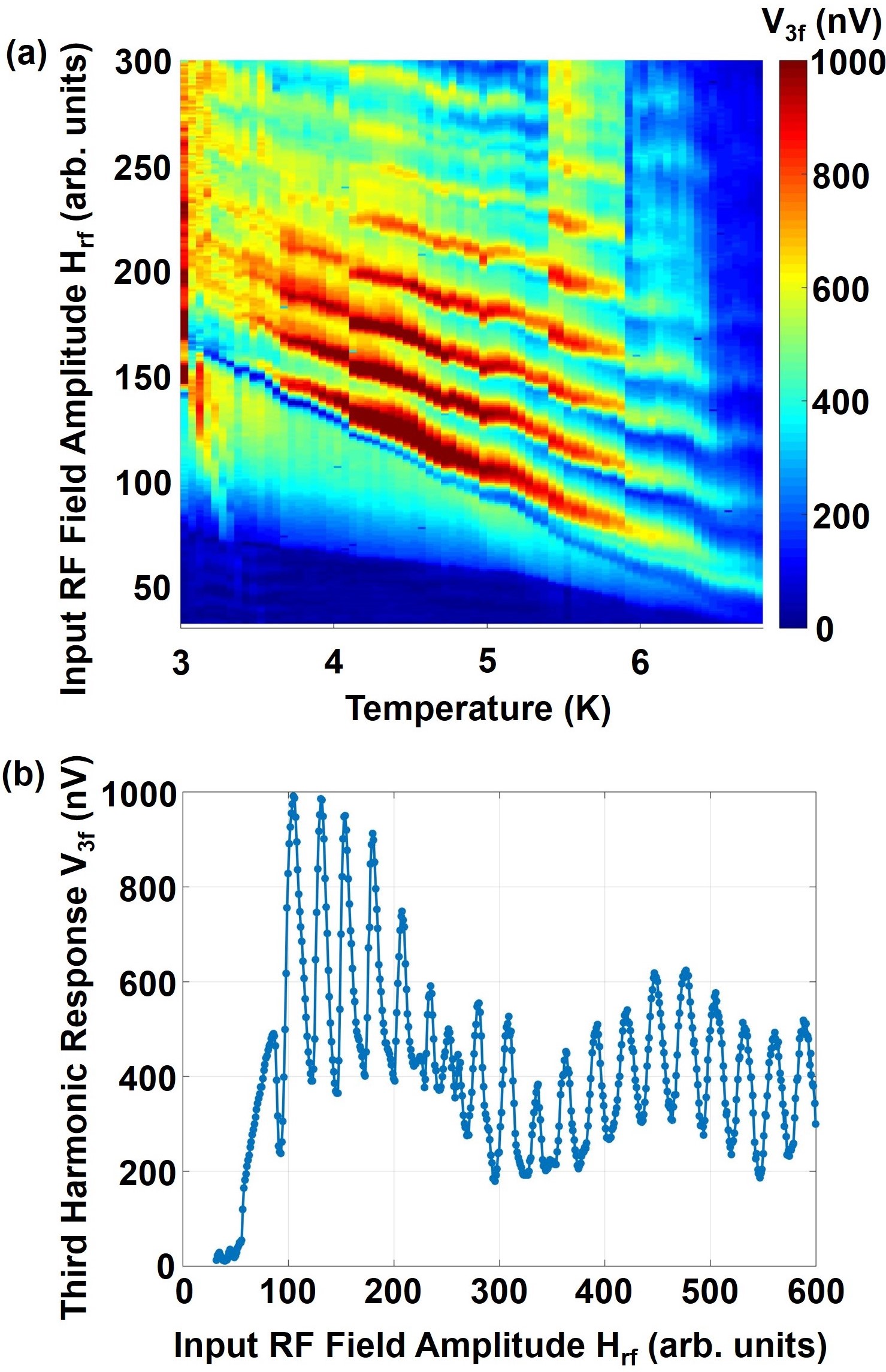}
    \caption{(a) Small-grain bulk Nb third-harmonic data vs temperature and applied rf field amplitude measured at 5.07 GHz rf input frequency. The sample is provided by G. Ciovati at Jefferson Lab. (b) Vertical line cut through this image ($V_{3f}$ vs $H_{rf}$) at a constant temperature of T=5.0 K.}
    \label{2ndBulkNb}
\end{figure}

Another set of bulk Nb samples is provided by the Ciovati group at Jefferson Lab. A high-purity fine-grain niobium sheet is cut into small samples by wire electrodischarge machining. Later, these samples are etched using buffered chemical polishing, heat treated at 600$\degree$C, and etched again. Afterward, the samples are nanopolished to obtain a surface with mirror-quality smoothness. Further details of sample preparation and characterization are available in ref. \cite{dhakal_effect_2018}. The data from this sample are very similar to the data obtained from large-grain bulk Nb sample. Fig.\ref{2ndBulkNb} shows representative nonlinear response data from sample F9 mentioned in \cite{dhakal_effect_2018}. \par

\begin{figure}
    \centering
    \includegraphics[width=0.45\textwidth]{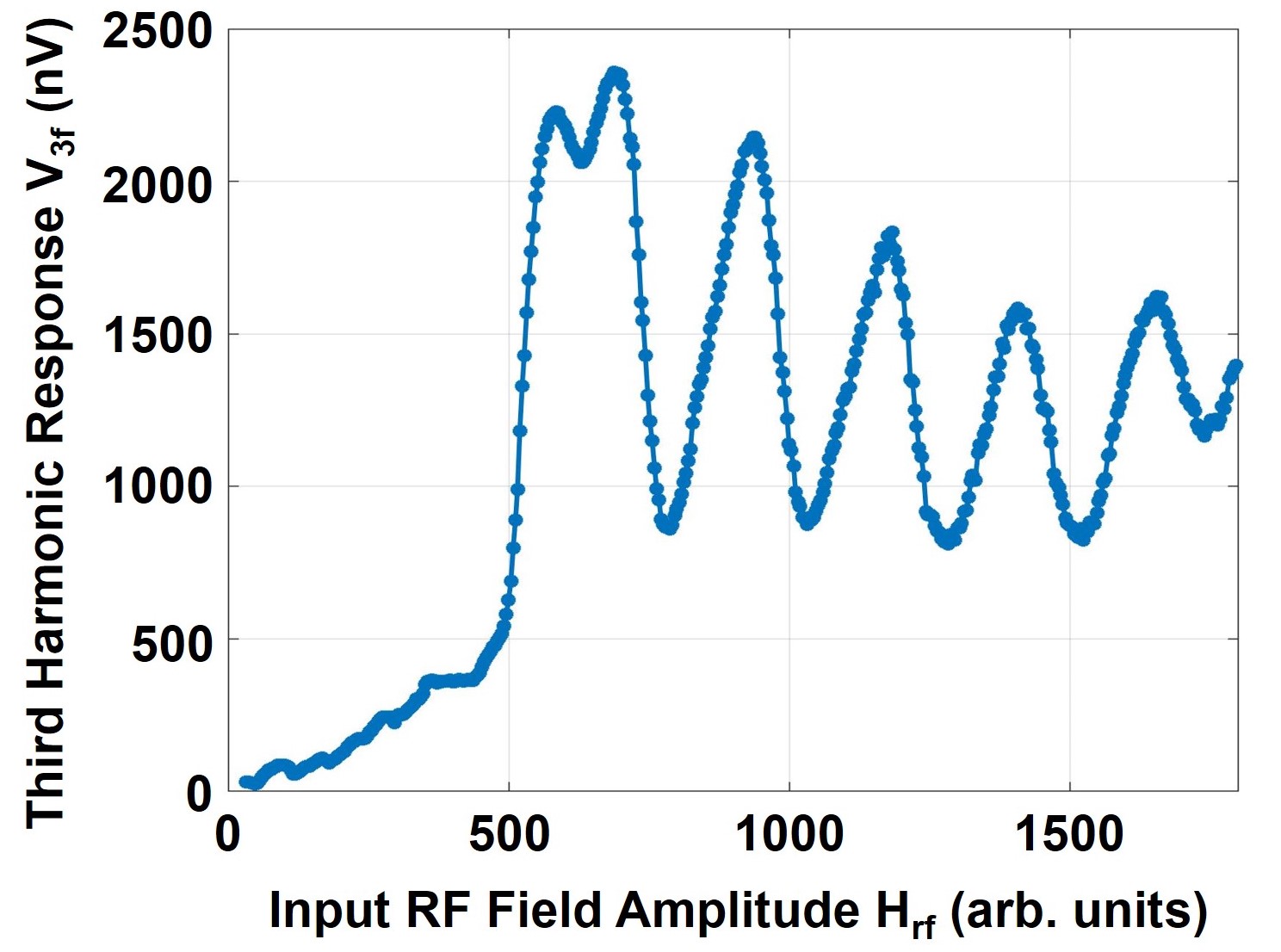}
    \caption{Third-harmonic response $V_{3f}$ vs input rf field amplitude $H_{rf}$ data at 7.5K and 4.855 GHz rf input frequency on Nb on Cu. The sample is provided by Oleg Malyshev's group in ASTeC.}
    \label{2ndFilmNb}
\end{figure}

A set of Nb on copper thin-film samples is provided by the Malyshev group at the Accelerator Science and Technology Centre (ASTeC). Four samples are deposited by high-power impulse magnetron sputtering at 700$\degree$C and various bias voltages on the Cu substrate \cite{OlegNbCu}. The plasma is driven with rectangular voltage pulses at a $200 Hz$ frequency and $100\mu s$ pulse duration. The peak discharge current for each pulse is approximately $40A$. All Nb film depositions are continued for 4 hours and the resulting films have a thickness of $1.4\mu m$. The residual resistivity ratio (RRR) is measured to be in the $42-54$ range, peaking for the film prepared at $0$ bias voltage. Fig.\ref{2ndFilmNb} shows representative nonlinear response data from one of those samples. A periodic response of $V_{3f}$ vs $H_{rf}$ is observed, similar to the other samples. \par

\begin{figure}
    \centering
    \includegraphics[width=0.45\textwidth]{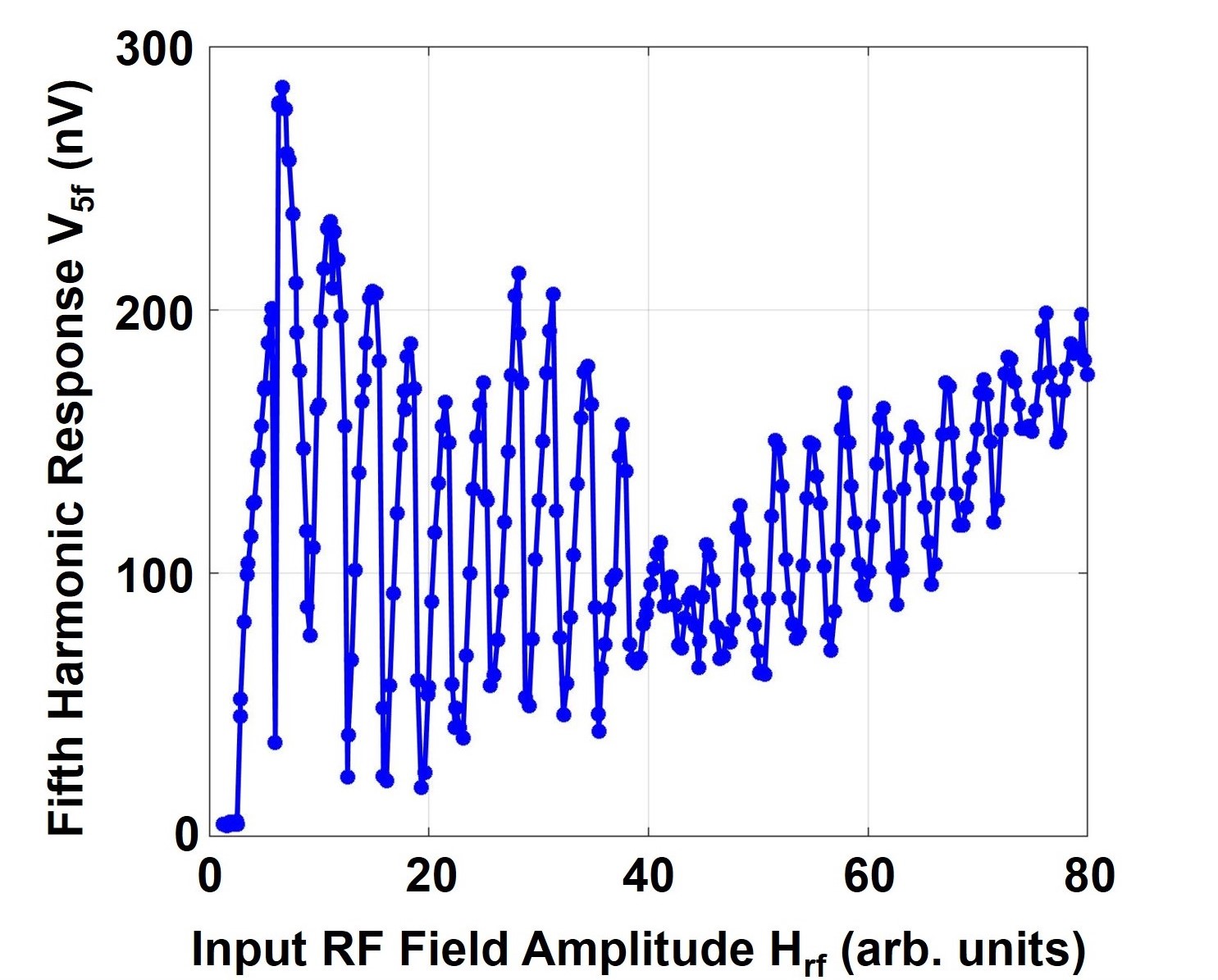}
    \caption{Fifth-harmonic response $V_{5f}$ vs input rf field amplitude $H_{rf}$ data at 4.2K and 1.361 GHz rf input frequency on Nb thin-film. The sample is provided by A. Valente-Feliciano's group at Jefferson Lab.}
    \label{AnneMarie}
\end{figure}

A set of Nb thin-films is provided by the Valente-Feliciano group at Jefferson Lab. The sample is coated using energetic condensation deposition in UHV via an electron cyclotron resonance (ECR) Nb ion source with a continuous ion energy of 184 eV \cite{AnneMarieThesis}. The sample has a thickness of 570 nm and is deposited on an $Al_2O_3$ substrate. The critical temperature is measured to be $T_c=9.18K$ and the $RRR$ is $31$. Fig. \ref{AnneMarie} shows the representative nonlinear response data from this sample. In this case, fifth-harmonic response $V_{5f}$ is measured and a very similar periodic response of $V_{5f}$ vs $H_{rf}$ is observed. A similar behavior of $V_{5f}$ is observed in the CERN film. We find that the driven RSJ model shows dips in $V_{5f}$ at the same $H_{rf}$ as $V_{3f}$. \par
The data from all these samples are very similar to those in Figs. 2,3 and 4 in the main text. The fact that such a wide variety of Nb samples produce a similar response indicates that this is a generic response of air-exposed Nb surfaces. \par

\bibliography{citeNbRCSJ}

\end{document}